# Using Animated Textures to Visualize Electromagnetic Fields and Energy Flow

John Belcher [a)]
Department of Physics
Massachusetts Institute of Technology

Carolann Koleci
Department of Physics
Worcester Polytechnic Institute

**Abstract**

Animated textures can be used to visualize the spatial structure and temporal evolution of vector fields at high spatial resolution. The animation requires two time-dependent vector fields. The first of these vector fields determines the spatial structure to be displayed. The second is a velocity field that determines the time evolution of the field lines of the first vector field. We illustrate this method with an example in magneto-quasi-statics, where the second velocity field is taken to be the **E**x**B** drift velocity of electric monopoles. This technique for displaying time-dependent electromagnetic fields has three pedagogical advantages: (1) the continuous nature of the representation underscores the action-by-contact nature of forces transmitted by fields; (2) the animated texture motion shows the direction of electromagnetic energy flow; and (3) the time-evolving field configuration enables insights into Maxwell stresses.

## I. Introduction

The representation of time dependent vector fields is a central problem in scientific visualization. There have been two advances in computer graphics since 1993 which combined have fundamentally changed the way time dependent vector fields can be visualized in two dimensions. The first of these was the introduction of the line integral convolution (LIC) method for showing the structure of vector fields at a resolution near that of the display, using textures generated by convolving the vector field with a grid of pixels of random brightness[1]. The second was the introduction of a method for the animation of a LIC by using a second velocity field to evolve the underlying grid of random pixels used to generate the LIC[2]. This latter method, dynamic line integral convolution (DLIC), produces an animated sequence of LIC images of the first field such that the time dependence of that field is evident from frame to frame by the inter-frame coherence in the LIC texture pattern.

In this paper we first discuss at a conceptual level how these two algorithms work. We then apply these methods to crossed electromagnetic fields, where the velocity field is taken to be the **E**x**B** drift velocity of electric or magnetic monopoles. We discuss the pedagogical utility of these electromagnetic animations in providing insight into electromagnetic phenomena. In addition to the obvious advantage of high spatial resolution, the DLIC method has three advantages: (1) it displays fields in a continuous way, which underscores the fact that the forces mediated by fields is via continuous contact; (2) it shows the direction of electromagnetic energy flow explicitly; and (3) the time-evolving field configuration enables insights into the Maxwell stresses. We provide an open source Java program that implements the DLIC algorithm described here, and which contains many examples of applications of this method to electromagnetism[3].

## II. Line Integral Convolution (LIC)

Most vector visualization algorithms use spatial structures to represent a vector field $\mathbf{F}(\mathbf{x})$. One of the most familiar is the field line representation, where a discrete set of curves are drawn which are everywhere parallel to the local field direction. Another is the vector field grid representation, where a set of icons on a fixed grid of spatial coordinates represents the field direction and perhaps magnitude at a given grid point. The use of field lines has the disadvantage that small scale structure in the field can be missed depending on the choice for the spatial distribution of the field lines. The vector field grid representation has a similar disadvantage in that the associated icons limit the spatial resolution because of the size of the icons and because of the spacing between icons needed for clarity. These two factors limit the usefulness of this representation in showing small scale structure in the field.

The LIC method avoids both of these problems by the use of a texture pattern to indicate the spatial structure of the field at close to the resolution of the display. To explain how the LIC algorithm works, first consider a constant field. Take a square array of *NxN* pixels of random brightness. We want to replace this white noise array with a textured array of the same dimension, where the texture pattern indicates the direction of the constant field, to within a sign. To do this, we process our *NxN* random array pixel by pixel to produce the new texture array, as follows. At any pixel 1 (see Figure 1), we average the brightness of the pixels along a line centered on pixel 1 and in the direction of the local field, for *n* pixels, $n << N$, and put this value in our new texture array at the same location as pixel 1 was in the initial array.

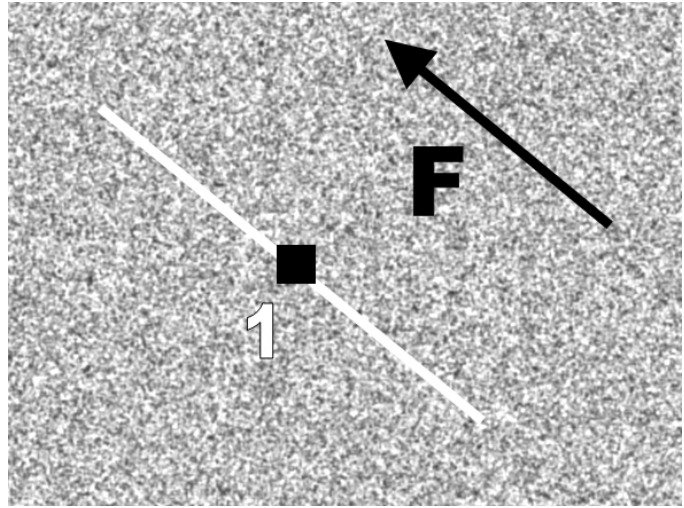

**Figure 1:** To produce a LIC image for the constant field **F**, we consider a pixel, for example pixel 1, and average the brightness of *n* pixels lying along a line parallel to **F** centered on pixel 1, as indicated by the while line.

      We now move to an adjacent new pixel and repeat this same process again (Figure 2). If we move parallel to the field to get to the new pixel, say pixel 2 in Figure 2, then the resulting average that we obtain at pixel 2 is almost the same as the average for pixel 1, because most of the pixels are the same. So the calculated brightness at pixel 2 is highly correlated with the brightness of pixel 1. If on the other hand we move perpendicular to the field to get to the new pixel, say pixel 3 in Figure 2, the resulting average is not correlated at all with the average at pixel 1, because none of the pixels whose brightness is being averaged are the same. This process produces a new array which has correlations in brightness along the field direction. Another way of saying this is that we have produced a texture pattern where the streaks in the texture are parallel to the field direction. The results of this process are shown in Figure 3.

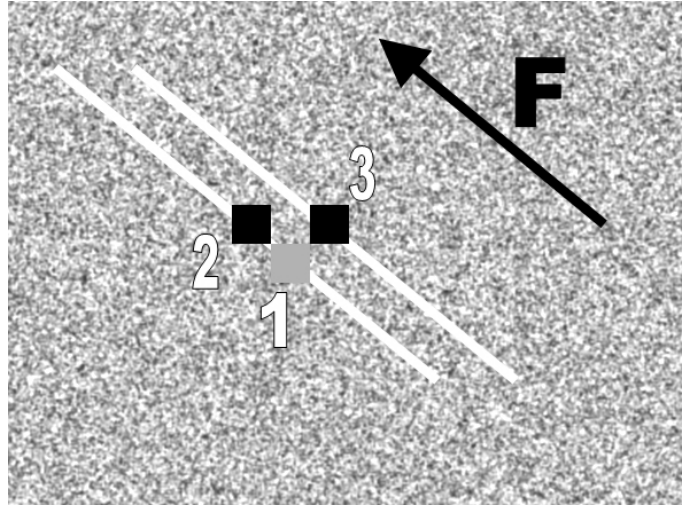

**Figure 2:** We calculate the brightness at pixels 2 and 3 by averaging over the brightness of the *n* pixels lying along the lines parallel to F centered on pixels 2 and 3, as indicated by the two white lines.

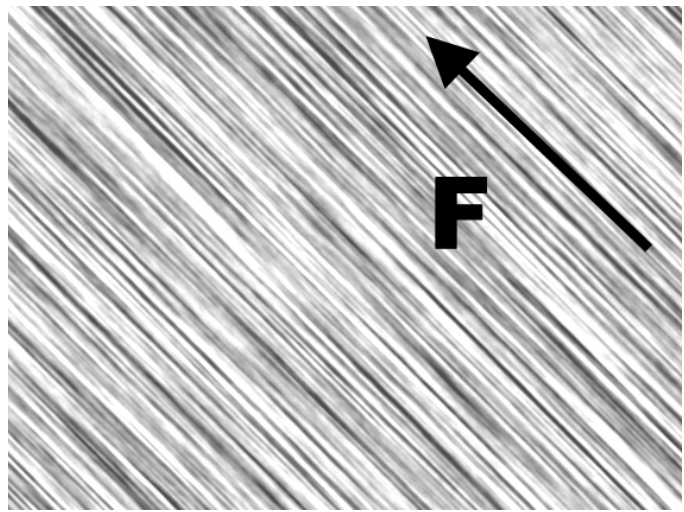

**Figure 3:** A LIC of a constant field, constructed in the manner described in the text.

Now consider the LIC procedure for a field that varies in space. If we simply follow the procedure described above and average the brightness of pixels along straight lines in space, where the direction of the straight line is determined by the local direction of the field at (for example) pixel 1, we would get a visual representation of the field but it would be inherently inaccurate, because we would be assuming that the local streamline can be reasonably approximated by a straight line along the entire *n* pixel averaging length. For locations where the local radius of curvature of a given field line is large compared to the *n* pixel length of the averaging line, this assumption is valid. However, if the local radius of curvature is comparable to or smaller than the length of *n* pixels along the averaging line, this assumption is no longer valid, and correlations in the texture pattern so generated will no longer show the details of structure of the field at this scale in a faithful manner.

To correct for this shortcoming, the Cabral and Leedom LIC[1] algorithm averages over *n* pixels along a line in space, but the averaging line is no longer a straight line. Instead it is the field line that passes through the point at which we are calculating the new texture value, that is pixel 1. That is, the texture pattern is convolved with the field structure along a line in space determined by the field lines (thus the name line integral convolution). This procedure retains the property that movement along the local field direction exhibits a high correlation in brightness values, but movement perpendicular to that direction exhibits little correlation, and this is true even in regions of high curvature.

Figure 4 shows a LIC for the magnetic field of a conducting ring falling toward a stationary magnetic dipole. Regions of high curvature occur near the two zeroes in the magnetic field strength just above the ring. The zeroes are distinguishable by the tilted X-like structure near them. For comparison we also draw four traditional field lines in the figure, to demonstrate that the correlations in the LIC are parallel to the field lines, even in regions of high curvature.

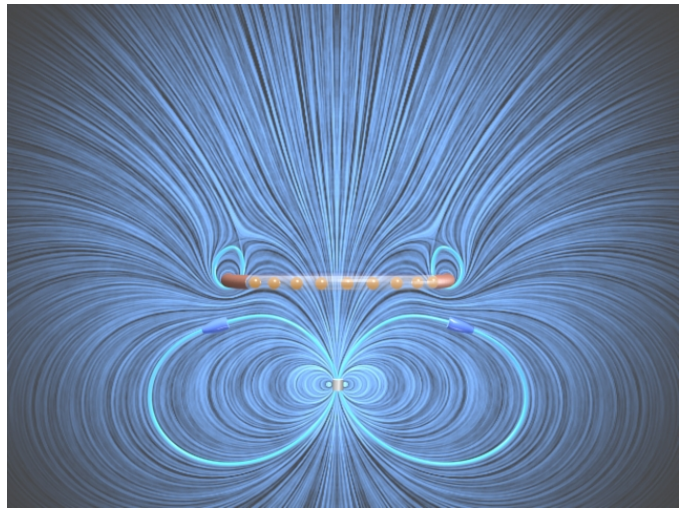

**Figure 4:** LIC of the magnetic field of a conducting ring as it falls toward a stationary magnetic dipole. The magnetic field shown includes both the field of the dipole and of the eddy currents induced in the ring.

### III. Dynamic Line Integral Convolution (DLIC)

The DLIC method extends the LIC algorithm described above to time-dependent fields. The vector field $\mathbf{F}(\mathbf{x},t)$ is allowed to vary with time, with the motion of its field lines described by a second velocity vector field, $\mathbf{D}(\mathbf{x},t)$. That is, at any time $t$ the field line passing through $\mathbf{x}$ at time $t$ is displaced in space at time $t + \Delta t$ to a new position $\mathbf{x} + \mathbf{D}(\mathbf{x},t)\,\Delta t$. To produce an animation, the DLIC algorithm originated by Sundquist[2] evolves the texture input used in LIC in a manner prescribed by the velocity field $\mathbf{D}$. That is, if $T(\mathbf{x},t)$ represents our random texture map discussed above, we evolve it with time according to

$$T(\mathbf{x},t+\Delta t) = T(\mathbf{x}-\mathbf{D}(\mathbf{x},t)\Delta t, t) \qquad (1)$$

Unfortunately, the texture pattern is typically stored as a discrete array of values on an ordered grid, and repeatedly evolving that array over time results in warping and a loss of detail because the velocity field $\mathbf{D}$ may have divergent or convergent regions, which will spread out or compress the location of the pixels in our texture. To avoid this problem, instead of evolving the input texture according to equation (1), the DLIC algorithm tracks a large number of particles of random intensity, roughly on the same order as the number of pixels in the original input texture. The particles move over time with a velocity given by $\mathbf{D}$, and the DLIC algorithm continuously monitors and adjusts their distribution to keep the level of detail roughly the same, by both consolidating and creating particles. At any instant of time for which we want to produce a frame of the animation, the texture at that time is generated by simply drawing all of the particles onto it. Once we have the texture for a given frame, the LIC method is applied to this texture to render the image of the field at that time.

Intuitively, since the particles that produce the input texture advect according to the motion field $\mathbf{D}$, the LIC convolution of a co-moving region of the field lines of $\mathbf{F}$ with the texture from one frame to the next samples the same part of the texture pattern, since the texture particles and field lines move in concert. Thus, the streaks in the LIC of $\mathbf{F}$ appear to move from one frame to the next according to the motion field $\mathbf{D}$. Each output image in the sequence will individually have the same properties as a static LIC rendering, but successive frames will have an inter-frame coherence that depicts the prescribed motion of the field lines.

### IV. Applications to Electromagnetism

The above discussion applies to the animation of any vector field for which we can specify a corresponding velocity field. We now turn specifically to electromagnetic vector fields and discuss an example of the construction of a DLIC in magneto-quasi-statics. We consider a situation in which the electric and magnetic fields are perpendicular. A conducting ring with mass $m$, radius $a$, resistance $R$ and self-inductance $L$ is located on the $z$-axis above a stationary permanent magnet with magnetic dipole moment vector $M_o\hat{\mathbf{z}}$. The normal to the ring is along the vertical z-axis, and the ring is constrained to move along that axis. The ring is released from rest at *t = 0*, and falls under gravity toward the conducting ring. Eddy currents arise in the ring because of the changing magnetic flux as the magnet falls toward the ring, and the sense of these currents will be such as to slow the ring.

Belcher and Olbert[4] argue that the magnetic field lines in this case evolve with a velocity field given by

$$\mathbf{D} = \frac{\mathbf{E} \times \mathbf{B}}{B^2} \qquad (2)$$

This velocity field represents the guiding center motion of a set of low energy electric monopoles initially arranged along any given magnetic field line, as those monopoles drift in the time-dependent electric and magnetic fields. The dynamics of the ring can be formulated mathematically in terms of three coupled ordinary differential equations for the position *Z(t)* of the ring, its velocity *V*(t), and the eddy current *I(t)* in the ring. The rate at which energy goes into Joule heating in the ring is given by

$$\frac{d}{dt}\left[\tfrac{1}{2}mV^2 + mgZ + \tfrac{1}{2}LI^2\right] = -I^2 R \qquad (3)$$

We consider the particular situation where the resistance of the ring (which in this model can have any value) is identically zero, and the mass of the ring is small enough (or the field of the magnet is large enough) so that the ring levitates above the magnet. We let the ring start out at a rest a distance *h* above the magnet. The ring begins to fall under gravity. When the ring reaches a distance of about *a* above the ring, its acceleration slows because of the increasing current in the ring. As the current increases, energy is stored in the magnetic field, and when the ring comes to rest, all of the initial gravitational potential of the ring is stored in the magnetic field (that is, in the $\tfrac{1}{2}LI^2$ term in equation (1)). That magnetic energy is then returned to the ring as it "bounces" and returns to its original position a distance *h* above the magnet. Since there is no dissipation in the system for our particular choice of *R* in this example (zero), this motion repeats indefinitely. Figure 4 shows one frame from the DLIC that visualizes the motion described above. The full animation is available on the web[3].

### V. Pedagogical Advantages

What are the advantages to the student to this way of representing electromagnetic phenomena? In the field line representation, students frequently ask what is "between" the field lines we choose to exhibit. The DLIC animation helps with this conceptual confusion in that it constructs a texture representation of the field that is continuous and exists at every point in space. This construction underscores one of Faraday's great insights. His concept of fields was developed to replace "action at a distance" with the notion of action by continuous contact. Objects that are not in direct contact (objects separated by apparently empty space) exert a force on one another through the presence of an intervening mechanism existing in the space between the objects, that is, the field. The force between two objects is transmitted by direct contact from the first object to the intervening field immediately surrounding that object, and then from one element of space to a neighboring element, in a continuous manner, until the force is transmitted to the region of space contiguous to the second object, and thus ultimately to the second object itself. That is, a material object generates a field and through that field influences its immediate neighborhood and ultimately the behavior of objects remote from its location.

Another advantage of the animation technique is the reinforcement of Faraday's insights into the connection between the shape and dynamics of electromagnetic fields, that is, the connection between their shape and the forces that they transmit. This is expressed mathematically by the Maxwell stress tensor, which depends only on the local field configuration and strength. As an example of this, consider the animation of the falling ring shown in Figure 4. As the ring moves downward, it is apparent in the animation that the magnetic field "texture" is compressed below the ring. This makes it intuitively plausible that the compressed field enables the transmission of an upward force to the moving ring as well as a downward force to the magnet.

We know physically (cf. equation (3)) that as the ring moves downward, there is a continual transfer of energy from the kinetic energy of the ring to the magneto-quasi-static energy of the magnetic field. The DLIC makes this manifest, because the overall appearance of the downward motion of the ring through the magnetic field is that of a ring being forced downward into a resisting physical medium, with stresses in the medium that develop due to this encroachment. Thus it is plausible to argue based on the animation that the energy of the downwardly moving ring is decreasing as more and more energy is stored in the magneto-quasi-static field, and conversely when the ring is rising. Moreover, because the texture motion is in the direction of the Poynting vector, we can explicitly see electromagnetic energy flowing away from the immediate vicinity of the ring into the surrounding field when the ring is falling and flowing back out of the surrounding field toward the immediate vicinity of the ring when it is rising.

All of these features make watching the DLIC animation a much more informative experience than viewing any single image of this situation, since the animation exhibits the actual flow directions of electromagnetic energy everywhere in space, as well as showing the field shape at any instant of time, which determine the properties of the local Maxwell stresses.

## VI.  Use in Instruction

Students often find the subject of electromagnetism to be esoteric and difficult to understand. Conceptually the student must first grasp what the concept of a field means. They must then understand how the interaction of material objects are mediated by the fields they generate. For this last task, they must understand the details of how fields are generated and how they change with time. The mathematics used in understanding these field properties is abstract and difficult to master[5]. All of these factors together lead to a substantial cognitive load (for a discussion of this in the context of, for example, Ampere's Law, sees Manogue et al.[6]).

One might argue that the use of graphical visualizations increases the cognitive load that students already combat in learning electromagnetism, and there is some risk of that. However, if used in a consistent fashion with a consistent format through out the subject exposition, we feel that the use of graphical visualization tools can be a positive part of the conceptual learning in this subject. Animated textures are only one way to represent fields, and should be only one of multiple representations. For one thing, the computational process used to animate the textures is calculationally intensive, and cannot be done in real time. Thus the student is reduced to passive viewing of movies of the animated textures, which has many disadvantages. In contrast, field lines or vector field representations of fields can be calculated rapidly enough that the student can interact with e.g. Java applications showing interacting and evolving charges and currents and their fields in real time. In this way students can see and interact with this process, and this has many advantages in terms of learning.

The use of multiple representations of the same phenomena has been the subject of considerable research in the physics education research literature[7]. The general consensus is that representations are important for student learning and those students who learn the material in an environment that uses more representations are less affected by the representational format of problem statements. Here is what we consider an idea combination of representations of electromagnetic fields: (1) single static diagrams showing field representations in all three formats mentioned above; (2) textured animations of dynamical situations illustrating how fields mediate the electromagnetic interaction of material objects; this could be for example the falling ring animation of Figure 4, or of two interacting charges, or of the interaction of two rings of current, and so on; (3) real time interactive applications containing time varying field lines or vector field grids, which can be influenced by student input, either by the dynamic change of some parameter in the application (i.e. the charge of a moving interacting charge) or by active intervention in the dynamics by clicking and dragging on a given charge or current element to see how the other charges or currents present then respond dynamically to that change; (4) the ability to pause such real time simulations at any point in time to create a single static LIC, which can be rendered in tens of seconds, so as to connect this representation to the more standard representations, as well as to show the field configuration at the resolution of the display at selected times.

Building visualizations with the properties described above is not technically challenging, and we have already developed examples of these kinds of visualizations for freshman physics courses at MIT[8], using the DLIC technique describe above, with Java 3D and Shockwave simulations for real time interaction . We are in the process of developing an entire suite of such electromagnetic visualizations to be incorporated in physics classrooms via lecture demos, recitation exercises, online homework modules, and other assignment venues. We are studying how effective this suite of visualizations is in demonstrating and conceptualizing fundamental physical phenomena in the realm of electromagnetism. We are interested in student gains both in introductory courses and in sophomore/junior level courses in electrodynamics, and are using these curricula materials both in MIT freshman courses and at WPI in freshman and upper level courses. The results of these studies will be the subject of future publications.

## VII. Summary

Animated textures can be used to visualize the spatial structure and temporal evolution of vector fields at high spatial resolution. This is particularly useful in visualizing time-dependent electromagnetic fields where the magnetic and electric fields are perpendicular, since the motion of the texture patterns can be used to indicate the direction of electromagnetic energy flow in these systems. We feel that the use of this visualization technique in addition to more traditional techniques for visualization of vector fields can be an important aid in student's conceptualization of the idea of fields. We are in the process of investigating this hypothesis quantitatively using guided instruction in conjunction with visualizations of the fields using this and other visualization techniques. The Java code and documentation for the creation of the animated textures described here are freely available[3].

**Acknowledgments:** This work is supported by NSF Grant #06185580, and a grant from the Davis Educational Foundation.